# Tuning Optical Properties of Self-Assembled Nanoparticle Network with External Optical Excitation

Zeynep Şenel,[1] Kutay İçöz,[1] Talha Erdem[1,*]

[1]Department of Electrical-Electronics Engineering, Abdullah Gül University, Erkilet Bul. 38080 Kayseri, Turkey

*Corresponding Author: Talha Erdem, erdem.talha@agu.edu.tr

ABSTRACT

DNA-driven self-assembly enables precise positioning of the colloidal nanoparticles owing to specific Watson-Crick interactions. Another important feature of this self-assembly method is its reversibility by controlling the temperature of the medium. In this work, we study the potential of another mechanism to control binding/unbinding process of the DNA-functionalized gold nanoparticles. We employ the laser radiation that can be absorbed by the gold nanoparticles to heat their network and disassociate it. Here, we show that we can actively control the optical properties of the nanoparticle network by an external optical excitation. We find out that by irradiating the structure with a green hand-held laser the total transmittance can increase by ~30% compared to the transmittance of the sample not irradiated by the laser. Similarly, the optical microscopy images indicate the transformation of the nanoparticle network from opaque to transparent while the nanoparticles formed a network again after the laser irradiation stopped. Our results prove that the optical excitation can be used to tailor the structure and thus the optical properties of the DNA-self-assembled nanoparticle networks.

**Introduction**

Fabrication of new generation complex, self-assembled structures in one, two, or three dimensions [1] using DNA-functionalized colloids [2] gives endless opportunities in various fields such as biology [3], photonics, [4,5] and nanodevices [6]. The introduction of DNA oligonucleotides attached to nanoparticles and the formation of lattices of nanoparticles in this way was first demonstrated in 1996 by Alivisatos et al. [7] and Mirkin et al. [8] These essential articles reported the crystal structures formed by the self-assembly of the gold nanoparticles mediated by the synthetic and short, single-stranded DNAs (ssDNA). They utilized the hydrogen bonds between the complementary ssDNA molecules attached to the nanoparticles. These nanoparticles were shown to stick to each other when the temperature decreases while they separate from each other at elevated temperatures. In the following years, several DNA-functionalization methods were studied and the relationships between the designed crystal structure and the number of DNA attached to the nanoparticles [9], the length of the DNA molecules are examined [10]. In recent years, the research efforts have moved towards utilizing the DNA-driven self-assembly for nanofabrication [11,12]. In these works, the nanostructures were first defined using the electron-beam lithography followed by the DNA-functionalization of these specific regions. Subsequently, DNA-functionalized nanoparticles were attached to these regions and metamaterials [11] and nanoparticle superlattices [12] were obtained. In addition to these efforts, optical interactions involving DNA-driven self-assembly has been extensively studied. For example, Simoncelli et al. employed polarization-dependent plasmonic heating with a femtosecond laser to selectively cleave the DNAs on the nanorods followed by attaching another DNA chain on the cleaved nanorod. Using this method, they managed to achieve nanoscale control on the molecular self-assembly [13]. In another work, Goodman et al. studied the release of the DNA molecules from the surface of the metal nanoparticles and showed that continuous wave lasers do not cause the DNA release whereas the pulsed laser excitation cleaves the DNA molecules on the gold nanoparticle surface [14]. Plasmonic heating of the nanoparticles has also been extensively employed in ultrafast photonic PCRs where efficient energy conversion from light to heat enables reduced nucleic acid amplification time [15,16]. Local heating with plasmonic particles also found applications in controlled drug release as shown by Song et al. [17] In this work, the cancer drug was loaded into Au nanoparticle assembled DNA hydrogel. By treating this hydrogel with light caused the release of the drug in a controlled manner.

Controlling the self-assembly of the metal nanoparticles with light has been studied in the literature with the help of additional molecules. For example, de Fazio et al. showed the reversible photoligation of the nanoparticles with DNA molecules and formed the superlattice of the nanoparticles when excited at 365 nm and unlocked this superlattice by exciting them at 312 nm [18]. Kanayama et al. employed photo-isomerization of an azobenzene moiety placed in close proximity of DNAs on gold nanoparticles to control their binding/unbinding process using light leading to a controlled color transition of the sample between violet and pink [19]. Using azobenzene modified DNAs to functionalize the nanoparticles, Zhu et al. modified the nanoparticle superlattice type between body-centered cubic and face-centered cubic owing to photoisomerization of the azobenzene [20].

Being inspired with these recent works, here we demonstrate the ability to control the binding/unbinding process of DNA-functionalized nanoparticles by irradiating them with a hand-held low-cost laser and employ the photothermal effect to tailor the optical features of the DNA-self-assembled nanoparticle networks in aqueous solutions without needing any other light-responsive chemical groups for the first time. Since the light emitted by the green laser is absorbed by the gold nanoparticles forming a network via DNA-DNA interactions, the nanoparticles heat their surroundings. Using optical microscopy, we showed that the applied laser light causes a reversible disassociation of the network owing to this optical heating effect. Furthermore, we examined the changes in the optical properties of nanoparticle network in account of this optical excitation causing the structural changes in the network. In this context, we observed that the laser excitation increased the transmittance of the nanoparticle network by ~30% in the visible regime. We believe that our results will pave the way for novel applications of DNA-driven self-assembly especially in actively controlling near-field interactions between different types of nanomaterials and all-solution-processed fabrication technologies.

**Experimental Methods**

All the chemicals used are purchased from Sigma Aldrich unless otherwise stated. The gold nanoparticles (Au NPs) were synthesized by reducing hydrogen tetrachloroaurate ($HAuCl_4$) and using a trisodium citrate solution following Ref [21].

The synthesized Au NPs were functionalized with thiol-modified single-stranded DNAs following the methods developed by Mirkin and co-workers [8,22–25]. In a typical experiment, the Au NPs are centrifuged for 1 h at 30,000 rcf to remove any residues left from the synthesis. Subsequently, the precipitated Au NPs are taken into a mixture of phosphate buffer (10 mM, pH=7.4) and sodium dodecyl sulfate (SDS, 0.015 w%) solution. Here, SDS molecules avoid the aggregation of the gold nanoparticles. Next, two complementary DNA thiol-functionalized DNA chains (obtained from Ella BioTech) were separately added to the gold nanoparticles. The single-stranded DNAs that we use have the following bases: (1) 5'-TTTTTTTTTTTTTTGGTGCTGCG-3', and (2) 5'-TTTTTTTTTTTTTTCGCAGCACC-3'. To increase the number of DNAs connected to each gold nanoparticle, NaCl solution in 0.015 w% SDS and 10 mM PB mixture is added stepwise such that the total salt concentration reaches 0.7 M within 3 h. Following the final salting step, the tubes are shaken overnight. The DNA-coated nanoparticles are then cleaned by centrifugation three times to remove unconnected single-stranded DNAs and the salt from the solution. The precipitated gold nanoparticles are redispersed in 0.015 w% SDS-containing PB solution in the first two centrifuges. After the third centrifugation process, no SDS was added to the DNA-coated nanoparticles. Next, about 50 μL of the nanoparticles that have complementary DNAs surrounding them are taken from each tube, mixed, and NaCl solution in PB is added to increase the final salt concentration to 100 mM. After hybridization, the color of the solution varies from red-pink to violet-black indicating the self-assembly of the nanoparticles and the formation of the nanoparticle networks whereas heating the mixture turned the color to its initial case.

Prior to optical characterizations, the heated samples are then loaded between two microscope slides, sealed with epoxy, and left for cooling. An Ocean Optics halogen light source connected to a fiber is used to illuminate the microscope slides at a normal angle (spot size: 0.2 cm$^2$). The transmitted light is collected using a fiber equipped Ocean Optics spectrometer. The transmission measurements are carried out first by recording the spectrum of the transmitted light through the microscope slides loaded with PB and then measuring the spectrum of the light transmitted through the same type of microscope slides filled with the DNA-functionalized Au NPs. The reported transmittance indicates the ratio of the measurement taken with Au NPs to the measurement taken with only PB. The effect of the external light is evaluated by continuously illuminating the sample with a green hand-held laser pointer (Yopigo ESO-2000, spot size: 0.35 cm$^2$) whose optical intensity variation was presented in Figure S1 in the Supporting Information. The transmission spectra are recorded before the laser illumination and every 10 min. after the laser radiation is applied. Microscopy images of the samples are recorded using a Nikon transmission optical microscope prior to laser irradiation and every 30 min. after the laser irradiation started. The electron microscope images are taken using a Zeiss Gemini scanning transmission electron microscope.

**Results and Discussion**

In this work, we investigate the effect of light on the DNA-driven self-assembly of nanoparticles and show the opportunities to manipulate the optical response of the nanoparticle network by light excitation. The temperature of the ambient medium is the main mechanism that tailors the binding and unbinding process of the nanoparticles that are connected to each other via complementary single-stranded DNA molecules. If the temperature is above a critical temperature called the melting temperature, the complementary ssDNAs remain separated leading to isolated nanoparticles within the aqueous medium. When the temperature decreases, the hydrogen bonds of the complementary DNAs on the nanoparticles will prevail and a network of nanoparticles will form. Here, we would like to use the light as a tool to control this process. The light that can be absorbed by the nanoparticles constituting the nanoparticle network will increase the temperature within their close proximity. As a result, the temperature will increase and dissolve the network of the nanoparticles if it rises above the melting temperature. Once the nanoparticle network dissolves, this should affect the intensity of the scattered light leading to changes in the optical properties of the network.

To test the applicability of this idea, we synthesized gold nanoparticles possessing a plasmon peak around 525 nm such that the light emitted by the commercially available, low-cost green lasers can be absorbed. Subsequently, we functionalized these nanoparticles with thiol-functionalized single-stranded DNAs. We then mixed the nanoparticles with complementary DNAs and formed a network utilizing DNA-DNA interactions. After heating the mixture above the melting temperature to dissolve the network, we quickly placed it between two microscope slides and sealed them with epoxy to avoid evaporation during the experiment. Before starting the optical experiments, we kept the mixture at a temperature below the melting temperature for a long time to guarantee the formation of the network. The scanning transmission electron microscope image in Figure 1(a) shows this network of nanoparticles. We found the size of the nanoparticles to be 22.3 ± 4.6 nm using this electron microscopy image. Based on Haiss et al. [26] and the UV-Vis measurements, the concentration of the nanoparticles was found to be 0.81 nM.

In our first test, we measured the transmittance of a broad-band white light through the sample while illuminating the sample with a green laser at an oblique angle as shown in Figure 1(b). Our results in Figure 1(c) show that the transmittance first decreases up to 60 min. and subsequently increases upon continuous laser illumination. At 80 min. of laser exposure, the transmittance reaches its maximum and afterwards we do not observe a significant variation in the transmittance levels. To make a better comparison, we analysed the change in the intensity of the transmitted light by applying Equation 1, where $T(t,\lambda)$ stands for the measured transmittance as a function of the laser illumination duration (t) and the wavelength ($\lambda$).

$$\%\Delta T(t,\lambda) = 100 \times \frac{T(t,\lambda)-T(0,\lambda)}{T(0,\lambda)} \qquad \text{(Equation 1)}$$

The transmittance data processed according to Equation 1 are presented in Figure 1(d). This analysis shows that the transmittance of the sample decreases within the first 60 min. to about 65% of the initial transmittance value ($\Delta T \approx -35\%$). After 60 min., the transmittance starts to increase again at all the wavelengths of interest. After 70 min. the transmittance increases beyond its initial level and at 80 min. it takes values ~30% higher than the level at t=0 ($\Delta T \approx 30\%$) and remains unchanged until the end of the experiment. The decrease in the transmittance within the first hour of laser exposure may be due to the large light-scattering clusters temporarily covering a larger area due to unbinding from larger features. Eventually, these large clusters disassociate into smaller clusters causing decreased scattering at longer laser exposures. Another point worth mentioning is that the variations in the transmittance occur over the whole spectral region of interest without a distinct wavelength dependence except a slightly stronger increase in transmittance in the blue region.

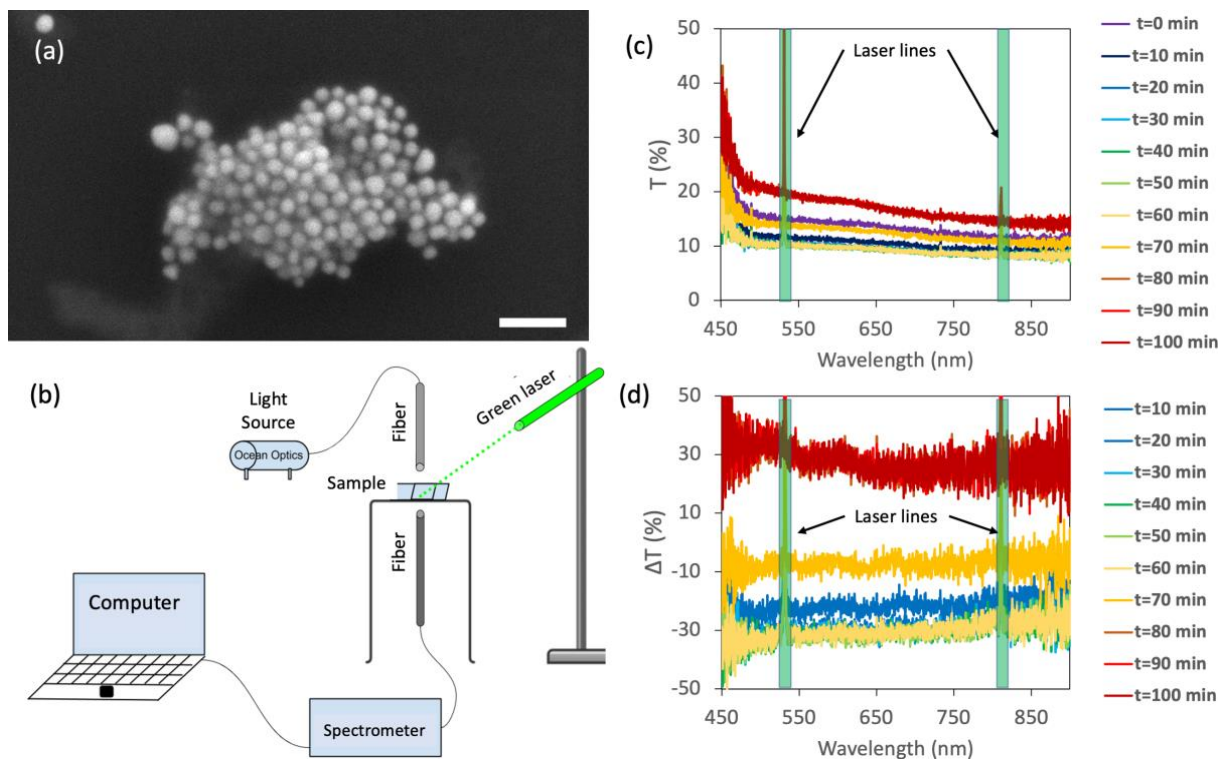

Figure 1. (a) Scanning transmission electron microscope image of the network of gold nanoparticles functionalized with complementary single-stranded DNAs. Scale bar: 100 nm. (b) Illustration of the transmittance measurement setup. (c) Transmittance (T) spectra of the DNA-self-assembled network of gold nanoparticles as a function of laser irradiation duration (*t*). (d) Change in the transmittance (ΔT) of the self-assembled nanoparticle network as a function of laser irradiation duration with respect to the transmittance of the sample before the laser exposure.

At this stage of our study, we designed a control experiment using a single type of DNA-functionalized gold nanoparticles. The aim of this experiment was to understand whether the observed increase in the transmission occurs due to the structural changes in the material network or another unpredicted effect related to our experimental setup and the gold nanoparticles. In our experiment, we placed these nanoparticles between microscope slides without hybridizing them with the nanoparticles functionalized with complementary DNAs. The measurements presented in Figure 2 do not show any obvious changes in the transmission levels regardless of the time of laser exposure when the sample is illuminated with a green laser. This shows that the observed changes in Figures 1(c) and (d) are directly related to the changes in the structure of the network.

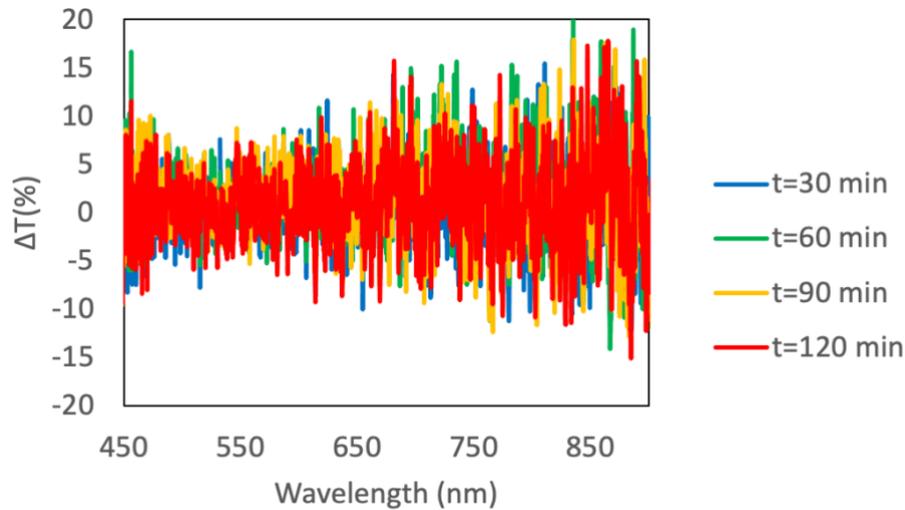

Figure 2. Change in the transmittance (ΔT) of the gold nanoparticles functionalized with a single-type of single-stranded DNAs as a function of laser irradiation duration with respect to the transmittance of the sample before the laser exposure.

To gain further understanding on the process, we measured the melting temperature of the DNA-functionalized nanoparticle mixture to be 51°C (Figure S2). Using an infrared temperature, upon laser exposure for 2h, we did not observe any increase beyond 32°C in the temperature of the solution containing the nanoparticle mixture. The temperature remaining below the melting temperature suggests that no bulk heating in the sample occurs due to the laser exposure. On the contrary, the laser light modifies the structure of the nanoparticle network owing to the laser light locally heating up the nanoparticles and their close proximity. Whether the separation of the nanoparticles is due to the separation of the connected DNA molecules on the nanoparticles or due to the disconnection of the ssDNAs from the nanoparticles is an important question. To test this, we exposed our sample with the laser at another time while measuring the transmittance (Figure S3). Due to the reorganization of the nanoparticles, we observed different absolute transmittance values than shown in Figure 1. However, the variation in the transmittance (ΔT) shows that an increase of the transmittance occurs similar to Figure 1 at the end of the experiment. Reproduction of the increased transmittance of the same sample shows that the binding-unbinding process of the nanoparticles is reversible process. This suggests that laser exposure causes the separation of the DNA molecules connecting the nanoparticles together rather than separating the DNA molecules from the surface of the gold nanoparticles.

Despite its limitations on detecting the changes of the clusters with small sizes, the optical microscopy still provides valuable information on how the structure evolves over time upon laser irradiation (Figure 3). Prior to laser irradiation, we see the broad dark regions inside the aqueous medium that indicates the clusters of gold nanoparticles. We did not observe any changes in the structures although we illuminate the nanoparticle network with white light. However, as shown in Figure 3, upon laser illumination we observed the bright spots opening inside the dark features showing that the laser light causes the nanoparticles to get separated from each other. To further evaluate whether the disassociation of the network is due to beam damage or due to unbinding of the nanoparticles owing to laser irradiation, we quickly cooled the sample at 4°C and observed the recovery of the network. The reversibility of the network shows that the laser beam does not cause a damage on the nanoparticle surface but locally heats the nanoparticle surroundings leading to unbinding of the nanoparticles.

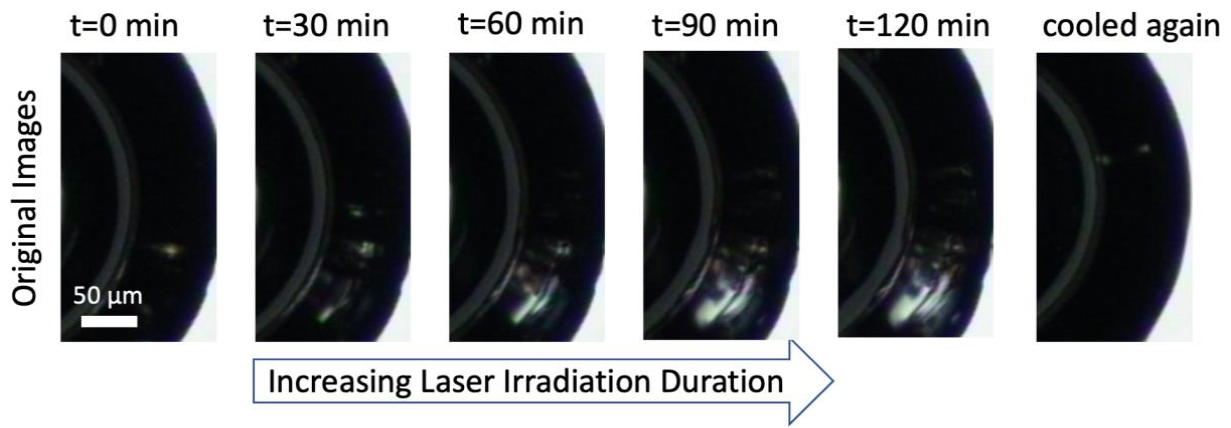

Figure 3. Optical microscopy images of a cluster of nanoparticles functionalized with complementary DNAs. As the sample is exposed to green laser light, the cluster of nanoparticles is observed to disassociate. When the sample is cooled again at 4°C (the right most image), we observe the disassociated region recovers indicating the reversibility of the process.

To retrieve a quantitative information out of the microscope images, we calculated the area covered by the white areas on the microscope images that correspond to the areas not covered by the nanoparticle network. As shown in Figure 4(a), we observe that the white areas on the image cover ~40% larger area at the end of the test compared to the initial white areas indicating that the nanoparticle network dissolves as a response to laser illumination. To compare this behaviour with the transmission measurements, we first integrated the transmittance presented in Figure 1(c) over the whole spectrum and plotted the integrated transmittance variation relative to the integrated transmission before the laser illumination starts (Figure 4(b)). This graph shows a qualitatively similar trend to Figure 4(a) both of which indicate increased transmission owing to the disassociation of the nanoparticle network.

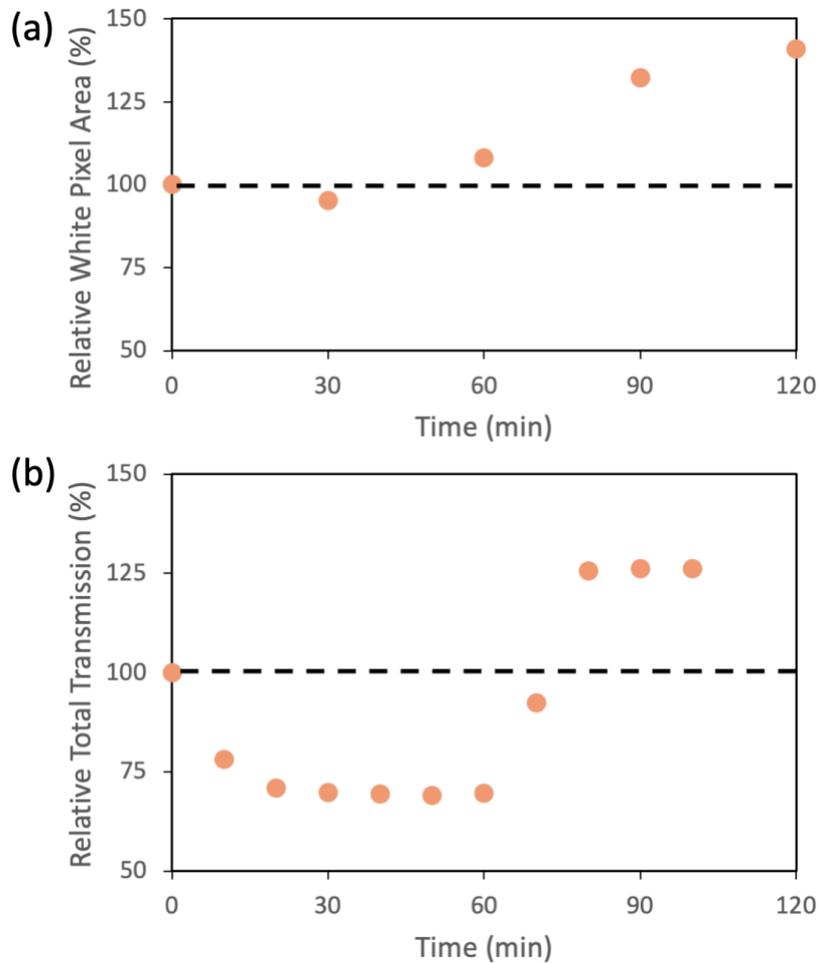

Figure 4. Time evolution of the disassociation of the self-assembled nanoparticle networks as calculated using optical microscopy and transmission spectroscopy. (a) Relative change of the area covered by white pixels on the microscope images corresponding to the areas not covered by the large nanoparticle clusters blocking the light as calculated using the optical microscopy images in Figure 3. The relative white areas at different time points are calculated with respect to the number of white pixels at t=0. (b) The time-evolution of the relative area under the transmittance spectrum as a function of the laser irradiation duration. The calculation is made using the transmittance information presented in Figure 1 by calculating the area under the spectrum between 450-900 nm. The data at different time points are normalized in accordance with the integrated transmittance at t=0.

**Conclusions**

The nanoparticles that possess complementary DNAs form networks of nanoparticles at low temperatures whereas this network is diminished when the temperature of the medium is increased. In this work, we showed that the laser irradiation can be used to control the binding and unbinding process of the DNA-functionalized nanoparticles inside a solution. As the laser light is absorbed by the nanoparticles, heating occurs in the close proximity of the nanoparticles and thus, the complementary DNAs that connect the colloidal particles separate from each other. As a result, the network of the particles disassociates as a response to the applied light. The reversibility of this process shows that the disassociation of the network does not occur due to beam damage to the nanoparticles but owing to local heating of the nanoparticle network by laser irradiation. We clearly observed the structural changes on the particle clusters upon laser irradiation using optical microscopy images. Furthermore, we showed that using externally applied light the optical transmission can be tailored. Our results enable controlling the structure and the optical properties of particles that are self-assembled via DNA-DNA interactions locally, paving the way for defining the structures of the particle networks locally in three dimensions using an external effect. Our approach can also be used to externally control the near-field interactions between the DNA-functionalized nanoparticles.

**Supplementary Material**

See supplementary material for the time-dependent laser intensity measurements (Figure S1), an additional time-dependent transmittance variation plot (Figure S2), and the melting temperature measurement (Figure S3).

**Acknowledgements**

TE is grateful to The Royal Society for the Newton International Fellowship Follow-on Funding Grant No. AL\201048. ZŞ acknowledges YÖK 100-2000 programme. We thank Mr. A. F. Yazici for taking electron microscopy images.

**Data Availability**

The data that support the findings of this study are available from the corresponding author upon reasonable request.